\documentclass[prc,twocolumn,showpacs,preprintnumbers,amsmath,amssymb]{revtex4-1}
\usepackage{graphicx}
\usepackage{bm}
\begin{document}
\title{Magnetization and collective excitations of a magnetic dipole fermion gas}
\author{Mitsuru Tohyama} 
\affiliation{
Kyorin University School of Medicine, Mitaka, Tokyo
181-8611, Japan}
\begin{abstract}
The ground states and collective excitations of trapped Fermion gases consisting of atoms with magnetic dipole moment are studied using
a time-dependent density-matrix 
approach. The advantages of the density-matrix approach are that one-body and two-body observables are directly calculated using one-body and 
two-body density matrices and that it has a clear relation to the Hartree-Fock (HF) and time-dependent HF theory. 
The HF calculations show the magnetization of the gases when the dipole-dipole interaction is strong. It is shown that
the tensor properties of the dipole-dipole interaction are revealed in the excitation modes associated with spin degrees of freedom.
\end{abstract}
\maketitle
\section{Introduction}
A degenerate Fermi gas of $^{161}$Dy has recently been achieved \cite{lu2} following the realization of Bose-Einstein condensation of $^{164}$Dy \cite{lu1}.
The dysprosium isotopes have large magnetic moments 10$\mu_B$ ($\mu_B$ being the Bohr magneton) and 
the progress of these experiments provides an opportunity to study exotic many-body physics with
magnetic dipolar moments.
Cold atomic systems with a synthetic spin-orbit coupling have also attracted strong experimental and theoretical interests \cite{lin,sau,yu,hu,sau}.
The dipole-dipole interaction is essentially a spin-orbit coupled interaction.
Sogo {\it et al.} \cite{sogo} and Li and Wu \cite{li} have recently demonstrated that in ultra cold dipolar Fermi gases the
dipole-dipole interaction can give rise to an instability toward spontaneous formation of a spin-orbit coupled phase.
They studied the properties of the spin-orbit couplings in infinite systems.
It is interesting to investigate how such phases with spin-orbit couplings are realized 
in trapped dipolar Fermion gases consisting of a finite number of atoms.
In this paper we study the ground states and collective excitations of a gas consisting of a small number of atoms with spin one half
using a time-dependent density-matrix approach (TDDMA) \cite{WC,GT}. Systems consisting of a small number of atoms have often been
used for theoretical investigations of dipolar Fermi gases \cite{Oster}
and may be realized in the array of microtraps or optical lattices as discussed
in Refs. \cite{Oster,Barberan,Popp}. 
The TDDMA consists of the coupled equations of motion for one-body and two-body density matrices. 
These equations are exact in the case of an $N=2$ system. The advantage of the TDDMA is that physical
observables are easily calculated using the one-body and two-body density matrices. Furthermore the TDDMA 
has a direct relation to the time-dependent Hartree-Fock approximation (TDHFA): 
Approximation of the two-body density matrix with anti-symmetrized products of the one-body density matrices in the TDDMA equation gives the TDHFA equation.
The TDDMA has recently been applied to polarized dipolar gases \cite{toh1,toh2} and a quantum dot \cite{toh3}.
The paper is organized as follows; the formulation is given in Sec. II, the results obtained for the ground state and the excited
states of an $N=2$ system are shown in Sec. III, the results for an $N=70$ system are presented in Sec. III, 
and Sec. IV is devoted to a summary.

\section{Formulation}
\subsection{Hamiltonian}
We consider a magnetic dipolar gas of fermions with spin one half, 
which is trapped in a spherically symmetric harmonic potential with
frequency $\omega$. The system is described by the Hamiltonian 
\begin{eqnarray}
H=\sum_\alpha\epsilon_\alpha a^\dag_\alpha a_\alpha
+\frac{1}{2}\sum_{\alpha\beta\alpha'\beta'}\langle\alpha\beta|v|\alpha'\beta'\rangle
a^\dag_{\alpha}a^\dag_\beta a_{\beta'}a_{\alpha'},
\label{totalH}
\end{eqnarray}
where $a^\dag_\alpha$ and $a_\alpha$ are the creation and annihilation operators of an atom at
a harmonic oscillator state $\alpha$ 
corresponding to the trapping potential $V(r)=m\omega^2r^2/2$ and
$\epsilon_\alpha=\omega(n+3/2)$ with $n=0,~1,~2,....$.
We use units such that $\hbar=1$ and assume that $\alpha$ contains the spin quantum number.
In Eq. (\ref{totalH}) $\langle\alpha\beta|v|\alpha'\beta'\rangle$ is the matrix element of 
a pure magnetic dipole-dipole interaction \cite{dipole}
\begin{eqnarray}
v(r)&=&-\frac{1}{r^3}\left(3({\bm d}_1\cdot\hat{\bm r})({\bm d}_2\cdot\hat{\bm r})
-{\bm d}_1\cdot{\bm d}_2\right)
\nonumber \\
&-&\frac{8\pi}{3}{\bm d}_1\cdot{\bm d}_2\delta^3({\bm r}),
\label{vdd}
\end{eqnarray}
where ${\bm d}$ is the magnetic dipole moment, ${\bm r}={\bm r}_1-{\bm r}_2$ and $\hat{\bm r}={\bm r}/r$.
The magnetic dipole moment for spin 1/2 is given by ${\bm d}=d{\bm \sigma}$ where ${\bm \sigma}$ is Pauli matrix.
In the case of completely polarized gases 
the second term on the right-hand side of Eq. (\ref{vdd}) can be neglected because the exchange term
cancels out the direct term. The contact term (the second term on the right-hand side of Eq. (\ref{vdd})) is usually
omitted in the study of dipolar gases. However, it is well-known that the contact term for the proton and electron magnetic 
dipole moments is essential to explain the hyperfine splitting of
a hydrogen atom. Therefore, in the following calculations we keep it as it is.
The effect of the contact interaction $g\delta^3({\bm r})$, which is usually additionally included in the study of cold atoms, is also 
considered in limited cases.

\subsection{$N=2$ system}
The TDDMA gives
the coupled equations of motion for the one-body density matrix (the occupation matrix) $n_{\alpha\alpha'}$
and the two-body density matrix $\rho_{\alpha\beta\alpha'\beta'}$.
These matrices are defined as
\begin{eqnarray}
n_{\alpha\alpha'}(t)&=&\langle\Phi(t)|a^\dag_{\alpha'} a_\alpha|\Phi(t)\rangle,
\\
\rho_{\alpha\beta\alpha'\beta'}(t)&=&\langle\Phi(t)|a^\dag_{\alpha'}a^\dag_{\beta'}
 a_{\beta}a_{\alpha}|\Phi(t)\rangle,
 \label{rho2}
\end{eqnarray}
where $|\Phi(t)\rangle$ is the time-dependent total wavefunction
$|\Phi(t)\rangle=\exp[-iHt] |\Phi(t=0)\rangle$.
The equations in the TDDMA are written as
\begin{eqnarray}
i \dot{n}_{\alpha\alpha'}&=&
(\epsilon_{\alpha}-\epsilon_{\alpha'}){n}_{\alpha\alpha'}
\nonumber \\
&+&\sum_{\lambda_1\lambda_2\lambda_3}
[\langle\alpha\lambda_1|v|\lambda_2\lambda_3\rangle \rho_{\lambda_2\lambda_3\alpha'\lambda_1}
\nonumber \\
&-&\rho_{\alpha\lambda_1\lambda_2\lambda_3}\langle\lambda_2\lambda_3|v|\alpha'\lambda_1\rangle],
\label{n2}
\end{eqnarray}
\begin{eqnarray}
i\dot{\rho}_{\alpha\beta\alpha'\beta'}&=&
(\epsilon_{\alpha}
+\epsilon_{\beta}
-\epsilon_{\alpha'}
-\epsilon_{\beta'}){\rho}_{\alpha\beta\alpha'\beta'}
\nonumber \\
&+&\sum_{\lambda_1\lambda_2}[
\langle\alpha\beta|v|\lambda_1\lambda_2\rangle\rho_{\lambda_1\lambda_2\alpha'\beta'}
\nonumber \\
&-&\langle\lambda_1\lambda_2|v|\alpha'\beta'\rangle\rho_{\alpha\beta\lambda_1\lambda_2}].
\label{N2C2}
\end{eqnarray}
Since there are no higher-level reduced density matrices in an $N=2$ system, these two equations are exact
if all elements of $n_{\alpha\alpha'}$
and $\rho_{\alpha\beta\alpha'\beta'}$ can be taken.
When the two-body density matrix in Eq. (\ref{n2}) is approximated by anti-symmetrized products of the occupation matrices,
Eq. (\ref{n2}) is equivalent to the equation in the TDHFA.

\subsection{$N\ge3$ system}
When the number of atoms is greater than two,
the equation of motion for the two-body density matrix is coupled to a three-body density-matrix
$\rho_{\alpha\beta\gamma\alpha'\beta'\gamma'}$:
\begin{eqnarray}
i\dot{\rho}_{\alpha\beta\alpha'\beta'}&=&
(\epsilon_{\alpha}
+\epsilon_{\beta}
-\epsilon_{\alpha'}
-\epsilon_{\beta'}){\rho}_{\alpha\beta\alpha'\beta'}
\nonumber \\
&+&\sum_{\lambda_1\lambda_2}[
\langle\alpha\beta|v|\lambda_1\lambda_2\rangle\rho_{\lambda_1\lambda_2\alpha'\beta'}
\nonumber \\
&-&\langle\lambda_1\lambda_2|v|\alpha'\beta'\rangle\rho_{\alpha\beta\lambda_1\lambda_2}]
\nonumber \\
&+&\sum_{\lambda_1\lambda_2\lambda_3}
[\langle\alpha\lambda_1|v|\lambda_2\lambda_3\rangle\rho_{\lambda_2\lambda_3\beta\alpha'\lambda_1\beta'}
\nonumber \\
&+&\langle\lambda_1\beta|v|\lambda_2\lambda_3\rangle\rho_{\lambda_2\lambda_3\alpha\alpha'\lambda_1\beta'}
\nonumber \\
&-&\langle\lambda_1\lambda_2|v|\alpha'\lambda_3\rangle\rho_{\alpha\lambda_3\beta\lambda_1\lambda_2\beta'}
\nonumber \\
&-&\langle\lambda_1\lambda_2|v|\lambda_3\beta'\rangle\rho_{\alpha\lambda_3\beta\lambda_1\lambda_2\alpha'}].
\label{N3C2}
\end{eqnarray} 
This coupled chain of equations of motion for reduced density matrices is known as the
Bogoliubov-Born-Green-Kirkwood-Yvon (BBGKY) hierarchy. 
The BBGKY hierarchy can be truncated by approximating the three-body density matrix with
the antisymmetrized products of the one-body and two-body density matrices
\cite{WC,GT}. As will be discussed below, however,
such a truncation is valid only in weakly interacting regimes.

\subsection{Ground State and Collective Excitations}
The ground state in the TDDMA is given as a stationary solution of the TDDM equations 
(Eqs. (\ref{n2}) and (\ref{N2C2})). 
We use the following adiabatic method to obtain a nearly stationary
solution \cite{adiabatic1}: Starting from a non-interacting spin-saturated configuration,
we solve Eqs. (\ref{n2}) and (\ref{N2C2}) gradually increasing the interaction 
$v({\bm r})\times t/T$. To suppress oscillating components which come from the mixing
of excited states, we must take large $T$.  We use
$T=2\pi/\omega\times 4$. For $t>T$ the interaction strength is fixed at $v({\bm r})$. We have checked the stability of the obtained 
ground state for $t>T$.
For strongly interacting regimes a spin-unsaturated deformed state becomes the ground state in the mean-field theory. 
In these regimes we perform
symmetry unrestricted Hartre-Fock (HF) calculations to obtain the HF ground state 
starting from a Slater determinant which breaks symmetries. 

We excite collective oscillations by introducing a time-dependent operator $\hat{Q}(t)$
to the total Hamiltonian Eq. (\ref{totalH}). In the case of a one-body excitation operator,
$\hat{Q}(t)$ is given by 
$k\sum_{\alpha\alpha'}\langle\alpha|Q|\alpha'\rangle
a^\dag_{\alpha}a_{\alpha'}\delta(t-T)$, where 
$k$ determines the oscillation amplitude. 
The initial conditions for the occupation matrix and the two-body density matrix at $t=T$ become such that
\begin{eqnarray}
n_{\alpha\alpha'}(T_+)=
\sum_{\lambda\lambda'}\langle\alpha|e^{-ikQ}|\lambda\rangle
n_{\lambda\lambda'}(T_-)\langle\lambda'|e^{ikQ}|\alpha'\rangle,
\label{n-init}
\end{eqnarray}
\begin{eqnarray}
\rho_{\alpha\beta\alpha'\beta'}(T_+)&=&
\sum_{\lambda_1\lambda_2\lambda_1'\lambda_2'}\langle\alpha|e^{-ikQ}|\lambda_1\rangle
\langle\beta|e^{-ikQ}|\lambda_2\rangle
\nonumber \\
&\times&\rho_{\lambda_1\lambda_2\lambda_1'\lambda_2'}(T_-)
\nonumber \\
&\times&\langle\lambda_1'|e^{ikQ}|\alpha'\rangle
\langle\lambda_2'|e^{ikQ}|\beta'\rangle,
\label{C-init}
\end{eqnarray}
where $T_-$ and $T_+$ indicate the times infinitesimally before and after $T$, respectively, and
$\langle\alpha|e^{ikQ}|\alpha'\rangle$ means
\begin{eqnarray}
\langle\alpha|e^{ikQ}|\alpha'\rangle&=&\delta_{\alpha\alpha'}-ik\langle\alpha|Q|\alpha'\rangle
\nonumber \\
&+&\frac{1}{2!}(ik)^2\sum_\lambda\langle\alpha|Q|\lambda\rangle\langle\lambda|Q|\alpha'\rangle+\cdot\cdot\cdot.
\end{eqnarray}
We study the collective modes in a small amplitude regime and, therefore,
expand Eqs. (\ref{n-init}) and (\ref{C-init}) up to second order of $k$.
The strength function $S(E)$ for an excitation operator $\hat{Q}$, which describes the distribution of
the transition strength, is calculated as \cite{toh1}
\begin{eqnarray}
S(E)=\frac{1}{k\pi}\int_0^\infty (q(t)-q(T))\sin Et'dt',
\label{strength}
\end{eqnarray}
where $q(t)=\langle \hat{Q}\rangle$ and $t'=t-T$. 
Since the integration in Eq. (\ref{strength}) is performed for a finite interval
in numerical calculations,
we multiply $q(t)-q(T)$ by a damping factor $\exp(-\Gamma t'/2)$ to suppress spurious oscillations
in $S(E)$. Since each discrete state gains an artificial width due to this damping factor,
$\Gamma$ must be smaller than experimental energy resolution.
We make a comparison of the TDDMA results with the TDHFA results.
The small amplitude limit of the TDHFA corresponds to the random-phase approximation (RPA) \cite{RS}.

\section{Results}
\subsection{Ground State}
\begin{figure} 
\begin{center} 
\includegraphics[height=6cm]{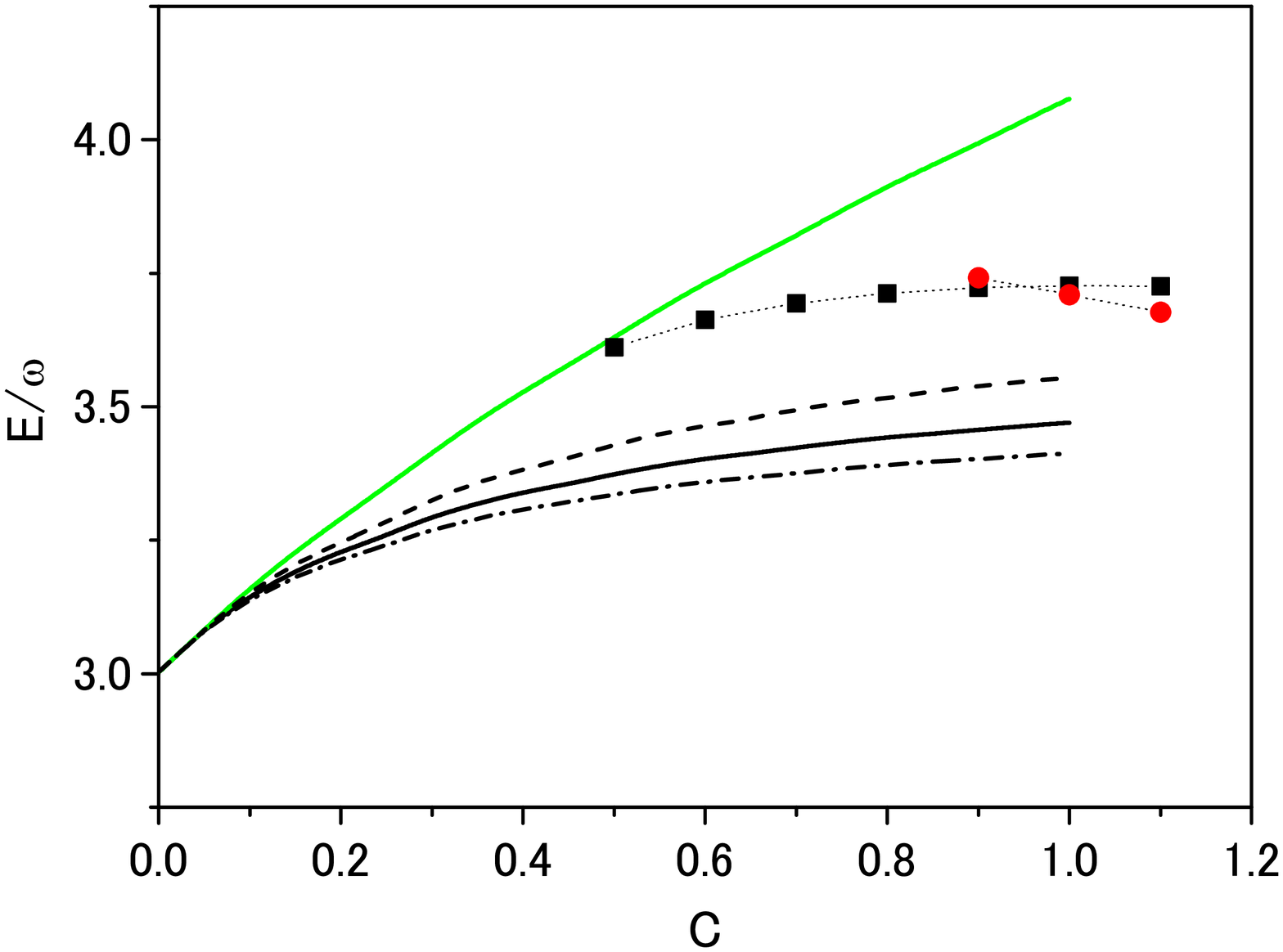}
\end{center}
\caption{(Color online) Ground-state energy as a function of $C=d^2/\omega\xi^3$ obtained in the TDDMA
for $N=2$ calculated with the single-particle states up
to the $2p-1f$ states (solid line). The dashed and dot-dashed lines show the TDDMA results calculated using the
single-particles states up to the $2s-1d$ and $3s-2d-1g$ states, respectively. The results in the TDHFA where spherical symmetry is imposed
are shown with the green (gray) solid line. The squares and circles denote
the results in the unrestricted HF approximation: The HF states denoted by the squares have a Rashba-like magnetization, while
those shown by the circles have magnetization in the $z$ direction.} 
\label{fig1} 
\end{figure}
\begin{figure} 
\begin{center} 
\includegraphics[height=6cm]{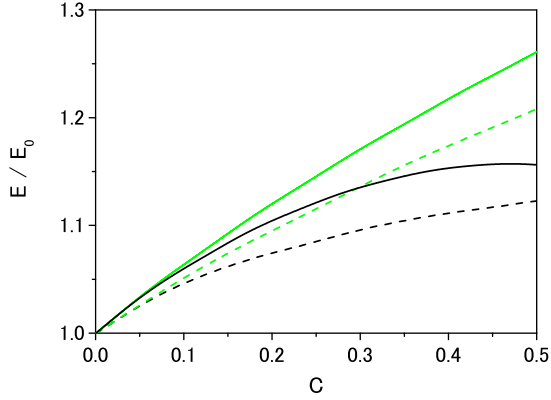}
\end{center}
\caption{(Color online) Ground-state energies calculated in the TDDMA (black solid line) and TDHFA (green (gray) solid line) for $N=8$.
The results for $N=2$ are also shown for comparison with the corresponding dashed lines.} 
\label{fig8} 
\end{figure}

 
First we consider an $N=2$ system, for which we can make a comparison of the results in the mean-field approaches and
the exact solutions given in the TDDMA.
As the starting ground state
we use a Slater determinant with a closed-shell configuration \cite{RS} where two atoms with spin up and down occupy the $1s$ state.
The number of $\rho_{\alpha\beta\alpha'\beta'}$ elements increases rapidly with increasing number of
the single-particle states, which makes it difficult to use a large number of the single-particle states.
For such a numerical reason we are forced to work with rather small configuration spaces but
this does not prevent from obtaining a semi-quantitative understanding of finite dipolar gases.
The ground-state energy calculated in the TDDMA for $N=2$ is show in Fig. \ref{fig1}
as a function of the parameter $C=d^2/\omega\xi^3$, where
$\xi$ is the oscillator length $\xi=\sqrt{1/m\omega}$. 
The dashed, solid and dot-dashed lines show the TDDMA results calculated using the single-particle states up to the $2s-1d$, $2p-1f$ and $3s-2d-1g$ states,
respectively. 
The range of $C$ considered in Fig. \ref{fig1} may be rather large 
for the current experimental situations \cite{lu2} : for example, $C$ 
for a gas of $^{161}$Dy trapped in a harmonic potential with $\omega=2\pi\times 500$Hz is about 0.2. A large value of $C$ may be
realized for a dipolar gas confined in a lattice \cite{lu2}.
The results in the TDHFA 
where spherical symmetry is imposed are shown with the green (gray) solid line.
Here the single-particle states up to the $2p-1f$ states are used.
In the case of the TDHFA calculations it is not so difficult to expand the single-particle space. From
the TDHFA calculations performed with the single-particle states up to the $3p-2f-1h$ states we estimate that the total HF energies
calculated with the single-particle states up to the $2p$ and $1f$ states explain 
$99.9$\% of the converged values. 
As shown in Fig. \ref{fig1}, the TDDMA results obtained using the single-particle states up to the $2p-1f$ states explain a substantial
part of the correlation energies though the TDDMA results are not completely converged. Therefore, in the following we mainly discuss the results obtained using 
the single-particle states up to the $2p-1f$ states.
The squares and circles denote
the results in the HF approximation without symmetry restriction, which will be discussed below.
The increase of the ground state energy with the increasing $C$ means that the interaction Eq. (\ref{vdd})
is repulsive. This is due to the contact term (the second term on the right-hand side of Eq. (\ref{vdd})). 
Note that the tensor part of Eq. (\ref{vdd}) alone cannot give any interaction energy when we start from
the non-interacting spin-symmetric ground state.
The difference between the TDDMA energy and the TDHFA energy is rather large, indicating the importance of the
ground-state correlations.
To investigate the effects of ground-state correlations in larger $N$ systems, we perform the TDDMA calculations for $N=8$
where a Slater determinant with the fully occupied $1s$ and $1p$ states (a closed-shell configuration) is used as the starting ground state.
We use the same single-particle states as those
used for $N=2$. The obtained results (black solid line) are
shown in Fig. \ref{fig8} as a function of $C$ and compared with the results of the spherical TDHFA calculations (green solid line). The results for $N=2$ are also shown for comparison.
The energy is normalized by the energy $E_0$ of the initial non-interacting state, which is $3\omega$ for $N=2$ and $18\omega$ for $N=8$.
As mentioned above the application of the TDDMA for $N\ge 3$ is limited to weakly interacting regimes $(C< 0.5)$. Figure \ref{fig8} suggests that the ground-state correlations are 
significant even in heavier systems.

\begin{figure} 
\begin{center} 
\includegraphics[height=6cm]{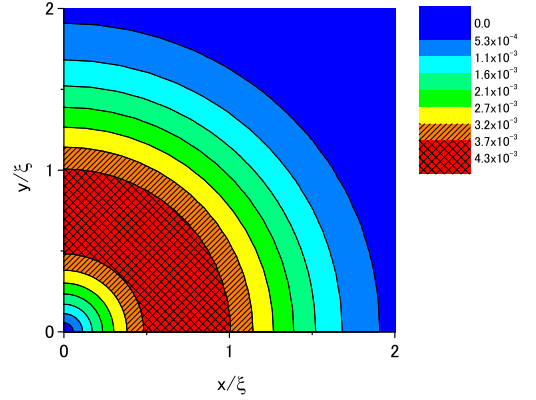}
\end{center}
\caption{(Color online) Contour plot of the distribution of the order parameter $\langle({\bm \sigma}\times {\bm r})_z\rangle$
in the $xy$ plane calculated in the unrestricted HFA for $N=2$ and $C=0.8$. The values of the order parameter
are given in arbitrary units.
The distribution has reflection
symmetry with respect to the $x$ and $y$ axes.} 
\label{fig2} 
\end{figure}

\begin{figure} 
\begin{center} 
\includegraphics[height=6cm]{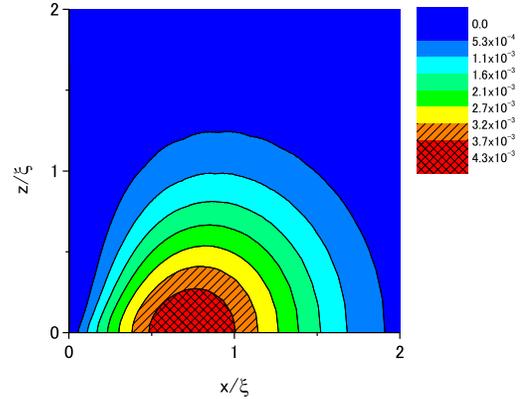}
\end{center}
\caption{(Color online) Contour plot of the distribution of $\langle({\bm \sigma}\times {\bm r})_z\rangle$
in the $xz$ plane calculated in the unrestricted HFA for $N=2$ and $C=0.8$.
The distribution has rotation
symmetry with respect to the $z$ axis and reflection symmetry with respect to the $x$ axis.} 
\label{fig3} 
\end{figure}

\begin{figure} 
\begin{center} 
\includegraphics[height=6cm]{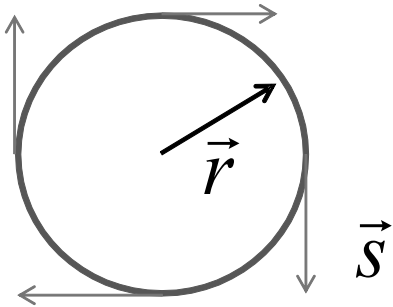}
\end{center}
\caption{Schematic picture for spin distribution of Rashba-like magnetization with $\langle\Phi_0|({\bm \sigma}\times {\bm r})_z|\Phi_0\rangle\neq 0$ .} 
\label{fig5} 
\end{figure}

\begin{figure} 
\begin{center} 
\includegraphics[height=6cm]{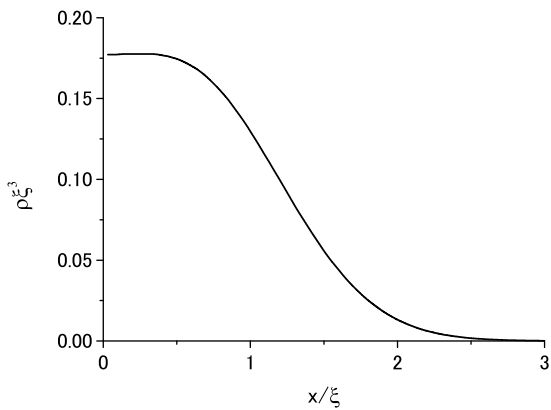}
\end{center}
\caption{Density distribution $\rho(x,0,0)$ as a function of $x$ in the unrestricted HFA for $N=2$ and $C=0.8$.
The density distribution is symmetric with respect to the origin.} 
\label{fig4} 
\end{figure}

Figure \ref{fig1}
shows that the breaking of spherical symmetry
gives a lower-energy solution in the HF approximation (HFA).
The HF ground states given by the squares are solutions with Rashba-like \cite{rashba} magnetization 
which are obtained starting from Slater determinants with $\langle\Phi_0|({\bm \sigma}\times {\bm r})_z|\Phi_0\rangle\neq0$. 
The circles in Fig. \ref{fig1} show the HF ground states where spins of the two atoms are completely polarized in the $z$ direction.
The tensor part of the dipole-dipole interaction is responsible for this completely polarized configuration because the 
contact term cancels out in such a configuration. 
In Figs. \ref{fig2} and \ref{fig3}
the distribution of the order parameter $\langle({\bm \sigma}\times {\bm r})_z\rangle$ which is given by
\begin{eqnarray}
\langle({\bm \sigma}\times {\bm r})_z\rangle&\equiv&
\langle\Phi_0|({\bm \sigma}\times {\bm r})_z\delta^3({\bm r}-{\bm r'})|\Phi_0\rangle
\nonumber \\
&=&
\sum_{\alpha\alpha'}({\bm \sigma}\times {\bm r})_zn_{\alpha\alpha'}\phi_\alpha({\bm r})\phi_{\alpha'}({\bm r})
\label{order}
\end{eqnarray}
is shown for the Rashba-like magnetized solution with $N=2$ and $C=0.8$. Here, $\phi_\alpha({\bm r})$ is the harmonic oscillator wavefunction.
Figures \ref{fig2} and \ref{fig3} show that the order parameter has a toroidal distribution.
The schematic picture of the spin distribution of this magnetized solution is shown in Fig. \ref{fig5} \cite{sogo}.
The density profile of the magnetized solution is shown in Fig. \ref{fig4}. 
The density distribution is spheroidally extended in the $xy$ direction (an oblate shape).
Figures \ref{fig2}, \ref{fig3} and \ref{fig4} show that the Rashba-like magnetization is realized mostly in the central part of the gas.

\begin{figure} 
\begin{center} 
\includegraphics[height=6cm]{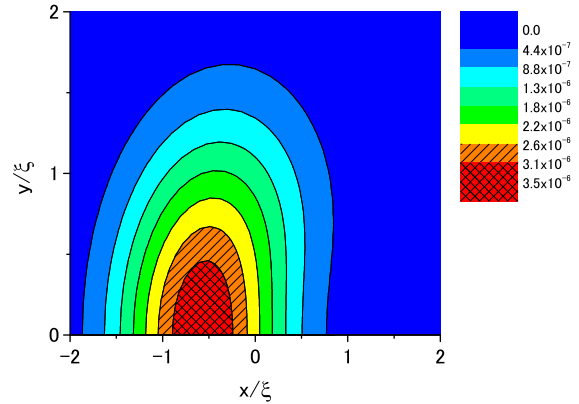}
\end{center}
\caption{(Color online) Contour plot of the two-body density distribution
$\rho({\bm r}\uparrow{\bm r'}\downarrow:{\bm r}\uparrow{\bm r'}\downarrow)$ (in  arbitrary units)
in the $xy$ plane calculated in the unrestricted HFA 
for $N=2$ and $C=0.8$, where ${\bm r'}=(1.25\xi,0)$. 
The distribution has reflection
symmetry with respect to the $x$ axis.} 
\label{fig.6} 
\end{figure}

\begin{figure} 
\begin{center} 
\includegraphics[height=6cm]{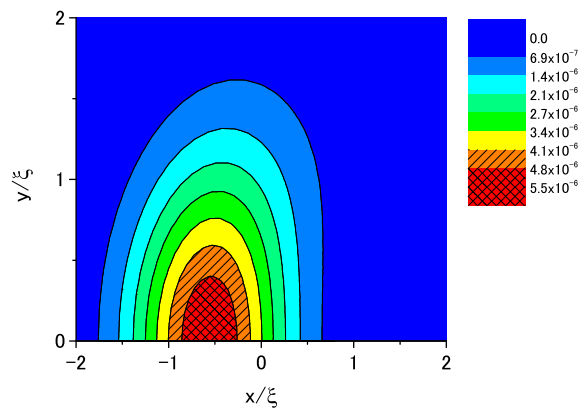}
\end{center}
\caption{(Color online) Same as Fig. \ref{fig.6} but calculated in the TDDMA.} 
\label{fig.7} 
\end{figure}

Since the ground-state calculation starts from the spin-saturated non-interacting configuration, the ground states in the TDDMA remain always spin-saturated and
have a spherically symmetric density distribution. In this case the order parameter, Eq. (\ref{order}), vanishes.
The TDDMA ground states are supposed to be a superposition of many configurations including magnetized and deformed ones.
To know the intrinsic structure of the TDDMA ground states, it is convenient to use 
the two-body density distribution $\rho({\bm r}s{\bm r'}s':{\bm r}s{\bm r'}s')$  which is given by the two-body density matrix as  
\begin{eqnarray}
\rho({\bm r}s{\bm r'}s':{\bm r}s{\bm r'}s')
&=&\sum_{\alpha\beta\alpha'\beta'}\rho_{\alpha(s)\beta(s')\alpha'(s)\beta'(s')}
\nonumber \\
&\times&\phi_\alpha({\bm r})\phi_\beta({\bm r'})\phi^*_{\alpha'}({\bm r})\phi^*_{\beta'}({\bm r'}).
\end{eqnarray}
This distribution gives the conditional
probability
to find an atom with spin $s$ at ${\bm r}$ when the other atom with spin $s'$ is located at
${\bm r'}$.
In the HFA the two-body density distribution is given as $\rho({\bm r}s{\bm r'}s':{\bm r}s{\bm r'}s')
=\rho({\bm r}s:{\bm r}s)\rho({\bm r'}s':{\bm r'}s')-\rho({\bm r}s:{\bm r'}s')\rho({\bm r'}s':{\bm r}s)$.
The contour plots of $\rho({\bm r}\uparrow{\bm r'}\downarrow:{\bm r}\uparrow{\bm r'}\downarrow)$ 
calculated in the unrestricted HFA and TDDMA for $N=2$ and $C=0.8$ are shown in Figs. \ref{fig.6} and \ref{fig.7}, respectively.
The position of ${\bm r'}$ is chosen at $(1.25\xi,0)$.
The two-body density distribution in the HFA is depleted in the region $x>0$ and enhanced in the region $x<0$,
which indicates $\rho({\bm r}\uparrow:{\bm r}\uparrow)\approx\rho({\bm r}\uparrow:{\bm r}\downarrow)$ 
and $\rho({\bm r}\uparrow:{\bm r'}\downarrow)\ll\rho({\bm r}\uparrow:{\bm r}\uparrow)$ for ${\bm r}\neq{\bm r'}$.
The two-body density distribution in the HFA is thus consistent with the magnetization shown in Fig. \ref{fig5}.
The two-body density distribution in the TDDMA is similar to that in the HFA. This suggests that the intrinsic structure in
the TDDMA ground state has the magnetization similar to the HFA ground state.

\begin{figure} 
\begin{center} 
\includegraphics[height=6cm]{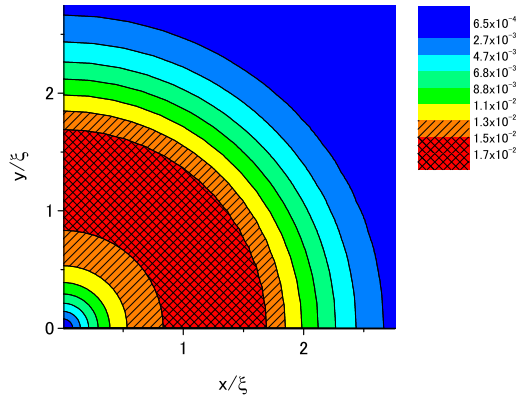}
\end{center}
\caption{(Color online) Same as Fig. \ref{fig2} but for $N=70$ and $C=0.8$.} 
\label{fig9} 
\end{figure}

\begin{figure} 
\begin{center} 
\includegraphics[height=6cm]{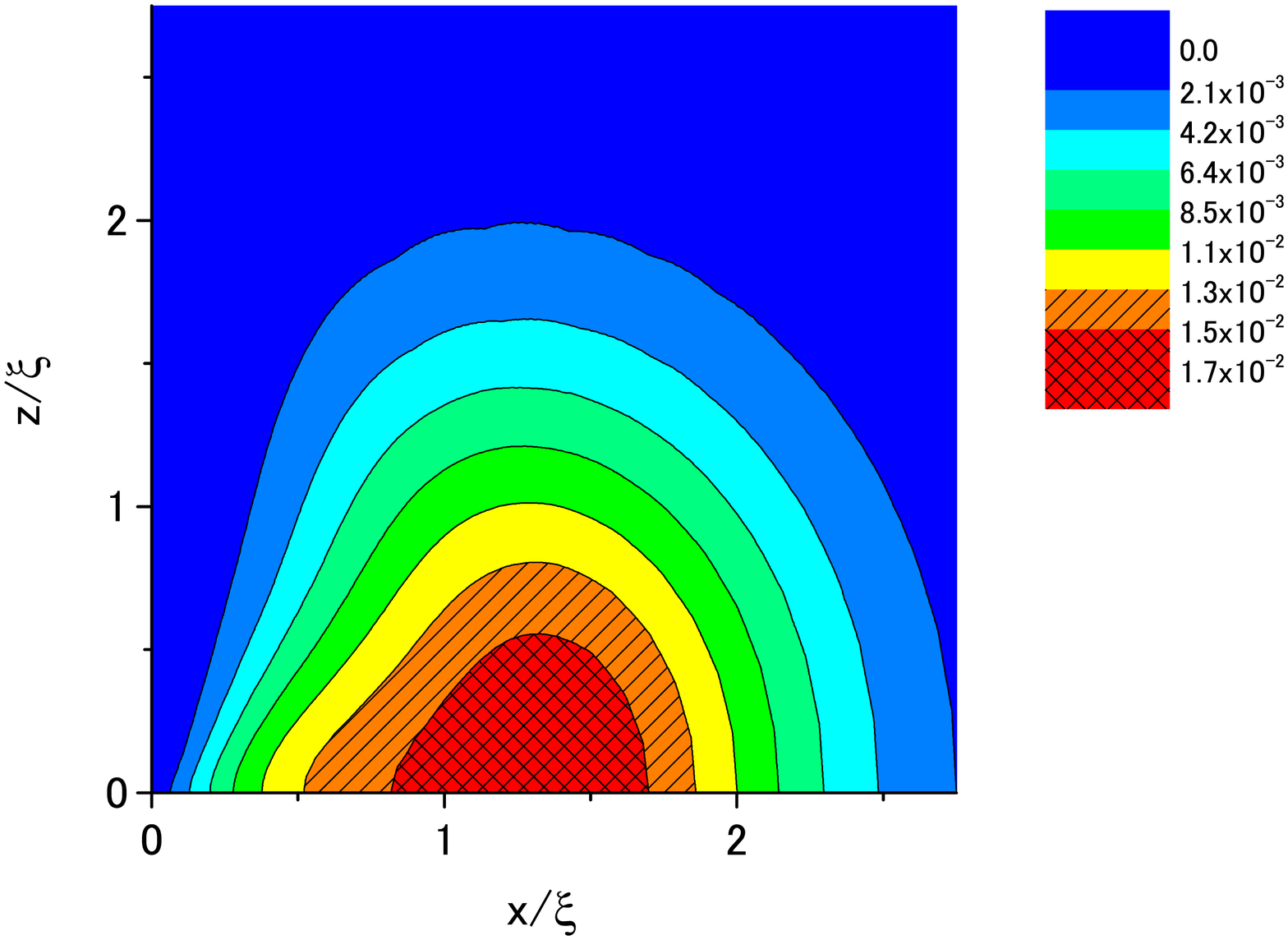}
\end{center}
\caption{(Color online) Same as Fig. \ref{fig3} but for $N=70$ and $C=0.8$.} 
\label{fig10} 
\end{figure}

\begin{figure} 
\begin{center} 
\includegraphics[height=6cm]{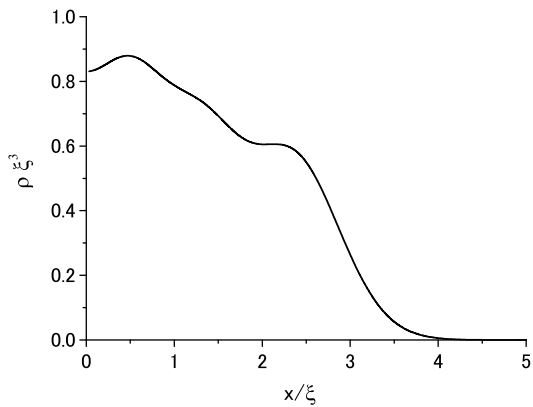}
\end{center}
\caption{Density distribution $\rho(x,0,0)$ as a function of $x$ in the unrestricted HFA for $N=70$ and $C=0.8$.} 
\label{fig11} 
\end{figure}

To study the magnetization in much heavier systems, 
we performed an unrestricted HFA calculation for $N=70$ and $C=0.8$ using the single-particle states
up to the $3p$, $2f$ and $1h$ states. 
The contour plots of $\langle({\bm \sigma}\times {\bm r})_z\rangle$ are shown in Figs. \ref{fig9} and \ref{fig10}. 
The density profile is
also shown in Fig. \ref{fig11}. It is thus found that a similar Rashba-like magnetization occurs in heavier systems.

\begin{figure} 
\begin{center} 
\includegraphics[height=6cm]{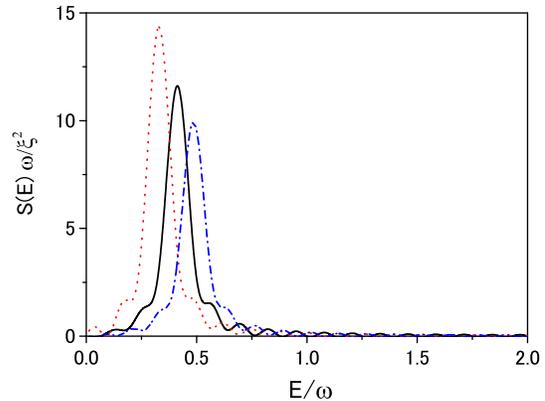}
\end{center}
\caption{(Color online) Strength functions for the Rashba-like mode calculated in the TDHFA for $N=2$ and $C=0.3$. The solid,
red dotted and blue dot-dashed lines show the results with $g/\omega\xi^3=0$, $1.8$ and $-1.8$, respectively.} 
\label{pmdlta} 
\end{figure}
 
It is pointed out in Refs. \cite{sogo} and \cite{li} that in infinite systems
the instability of the spin symmetric HF ground state against the spin monopole and spin quadrupole modes
occurs faster than the spin-orbit mode associated with $({\bm \sigma}\times{\bm r})_z$. As shown below,
we found that in the case of a trapped gas considered here, which is spherical symmetric and spin-saturated, the instability against
the spin monopole and spin quadrupole modes
occur for stronger dipole-dipole interaction ($C> 1$) than the $({\bm \sigma}\times{\bm r})_z$
mode. This is because the monopole and quadrupole modes should overcome $2\omega$ excitation energy in such trapped gases with
closed-shell configurations.
Trapped gases with open-shell configurations may have instabilities similar to infinite systems, which is an interesting subject of future study.

In order to investigate the effect of the contact interaction $g\delta^3({\bm r})$ on the instability of the Rashba-like mode, 
we calculated the strength function for the excitation operator $({\bm \sigma}\times{\bm r})_z$ in the TDHFA.
The obtained strength functions for the Rashba-like mode for $N=2$ and $C=0.3$
are shown in Fig. \ref{pmdlta} where $g\delta^3({\bm r})$ with $g/\omega\xi^3=0$ (solid line), $1.8$ (red dotted line) and $-1.8$ (blue dot-dashed line) are used.
Figure shows that the repulsive contact interaction makes the Rashba-like mode soft.
In fact we found that the simple repulsive contact interaction $g\delta^3({\bm r})$ alone can also give a Rashba-like magnetization 
when it is sufficiently strong ($g/\omega\xi^3>18$). 
The results for infinite systems \cite{sogo} also show 
that spin modes become unstable for a strongly repulsive contact interaction.

\subsection{Collective Excitations}
\subsubsection{Quadrupole Modes}
\begin{figure} 
\begin{center} 
\includegraphics[height=6cm]{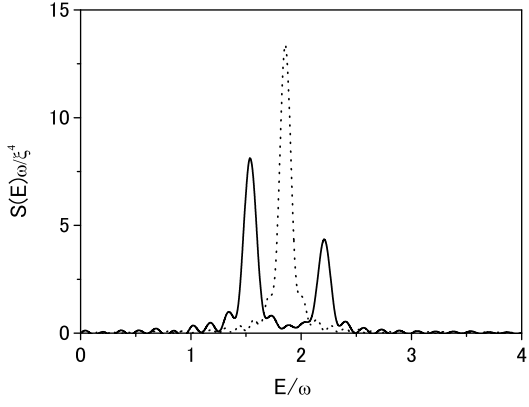}
\end{center}
\caption{Strength function for the quadrupole mode calculated in the TDDMA for $N=2$ and $C=1$ (solid line). 
The result in the TDHFA where the spherically symmetric HF ground state is used is shown with the dotted line.
The artificial width used is $\Gamma/\omega=0.1$} 
\label{fig12} 
\end{figure}

The strength function for the quadrupole mode calculated in the TDDMA (solid line) for $N=2$ and $C=1$ is shown in Fig. \ref{fig12}.
The excitation operator used is $(z^2-(x^2+y^2)/2)$. An excited mode is classified by the orbital angular momentum $L$, the total spin $S$ and
the total angular momentum $J$. Its parity $P$ is given by $P=(-1)^L$. The mode excited by $(z^2-(x^2+y^2)/2)$ has $L=2$, $S=0$ and $J^P=2^+$. 
The result in the TDHFA is shown with the dotted line. 
In the TDHFA calculation we used the spherically symmetric HF ground state so that the excited modes have good 
quantum numbers as do the results in the TDDMA.
The artificial width used is $\Gamma/\omega=0.1$. The TDDMA result is quite different from the TDHFA result which shows a single peak.
The split of the strength in the TDDMA is considered to be due to the decoupling of the quadrupole mode and two-phonon states.
The candidate of the two-phonon states that have $E\approx2\omega$ is the double Kohn mode.
The single Kohn mode is the center-of-mass motion which can be excited by the operator $z$ and it is well-known \cite{Kohn,Brey,Dobson} 
that the Kohn mode has excitation energy $\omega$
for any interaction with translational invariance.
We numerically confirmed this property.
Since the excitation operator for the double Kohn mode includes a one-body part such that
\begin{eqnarray}
\left(\sum_{\alpha\alpha'}\langle\alpha|z|\alpha'\rangle
a^\dag_{\alpha}a_{\alpha'}\right)^2&=&\sum_{\alpha\alpha'}\langle\alpha|z^2|\alpha'\rangle
a^\dag_{\alpha}a_{\alpha'}
\nonumber \\
&+&\sum_{\alpha\beta\alpha'\beta'}\langle\alpha|z|\alpha'\rangle\langle\beta|z|\beta'\rangle
\nonumber \\
&\times&a^\dag_{\alpha}a^\dag_{\beta}a_{\beta'}a_{\alpha'},
\end{eqnarray}
the double Kohn mode can be excited by the one-body operator $(z^2-(x^2+y^2)/2)$. 
It is pointed out in Ref.\cite{ts04} that the excitation 
energy of the double Kohn mode 
should be $E=2\omega$ for any interaction with translational invariance. 
The state which presumably consists of the double Kohn mode appears slightly above $2\omega$. 
Such a deviation may be due to the truncation of the single-particle space which makes it difficult to properly describe the two phonon states.
A clear splitting of the double Kohn mode with $E=2\omega$ is seen in the TDDMA calculations for the monopole and quadrupole excitations of a two-dimensional
quantum dot with $N=2$ \cite{toh3},
where a larger single-particle space can be taken.
\begin{figure} 
\begin{center} 
\includegraphics[height=6cm]{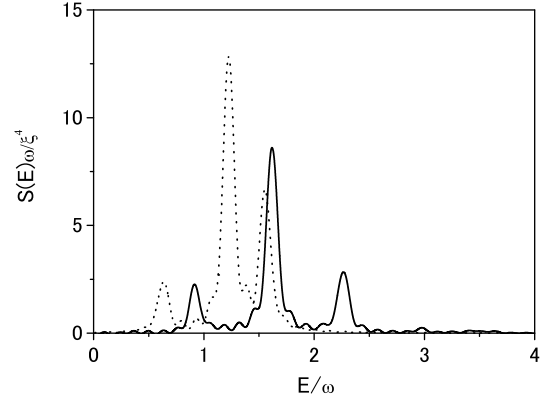}
\end{center}
\caption{Same as Fig. \ref{fig12} but for the spin quadrupole mode excited by the operator $\sigma_z (z^2-(x^2+y^2)/2)$.} 
\label{fig13} 
\end{figure} 

The strength functions calculated for the spin quadrupole mode are shown in Fig. \ref{fig13} for $N=2$ and $C=1$.
The excitation operator used is $\sigma_z (z^2-(x^2+y^2)/2)$ which can excite states with $L=2$, $S=1$ and $J^P=1^+$ and $3^+$.
The dipole-dipole interaction is 
strongly attractive in the particle - hole channel for the spin quadrupole modes \cite{sogo,li}.
Figure \ref{fig13} shows that the effects of the ground-state correlations 
strongly reduce the particle - hole correlations. Comparing with the spin monopole mode excited by the operator $\sigma_zr^2$ which also excites states with
$J^P=1^+$, we found that
the lowest and highest-energy states at $E/\omega=0.9$ and $2.3$ in Fig. \ref{fig13} calculated in the TDDMA 
have $J^P=1^+$ while the largest peak at $E/\omega=1.6$ corresponds to the $3^+$ state.
Since the lowest $1^+$ state is strongly excited by the operator $\sigma_zr^2$, its main component is considered to be $L=0$, $S=1$ and $J^P=1^+$.
The spin quadrupole mode with $L=2$, $S=1$ and $J^P=2^+$ comes between the lowest $1^+$ state and the $3^+$ state as a single peak, though it is not shown in Fig. \ref{fig13}.
We found that 
the simple contact interactions of the form $g\delta^3({\bm r}_1-{\bm r}_2)$ or ${\bm d}_1\cdot{\bm d}_1g'\delta^3({\bm r}_1-{\bm r}_2)$
gives a single peak for the spin quadrupole modes. Therefore, the splitting of the spin quadrupole modes depending on $J$ is caused by the tensor part
of the dipole-dipole interaction (the first term on the right-hand side of Eq. (\ref{vdd})).
 
\subsubsection{Spin Dipole Modes}
\begin{figure} 
\begin{center} 
\includegraphics[height=6cm]{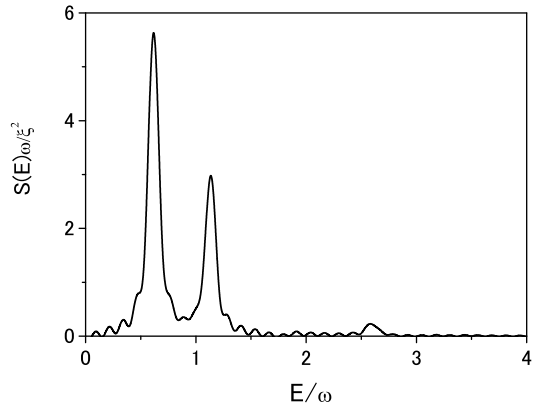}
\end{center}
\caption{Strength function for the spin dipole mode calculated in the TDDMA.} 
\label{fig14} 
\end{figure}

\begin{figure} 
\begin{center} 
\includegraphics[height=6cm]{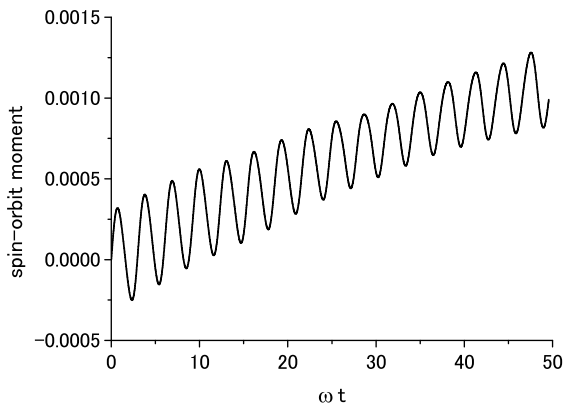}
\end{center}
\caption{Time evolution of the spin-orbit moment $\langle\Phi_0|({\bm \sigma}\times {\bm r})_z|\Phi_0\rangle$ (in arbitrary units) calculated in the TDDMA.} 
\label{fig15} 
\end{figure}

Finally we show the result for the spin dipole mode excited by the operator
$\sigma_z z$ which can excite states with $J^P=0^-$ and $2^-$. 
The strength function for the spin dipole mode calculated in the TDDMA 
is shown in Fig. \ref{fig14} for $N=2$ and $C=0.8$. Figure \ref{fig14} 
indicates that the spin dipole mode becomes quite soft for large interaction strength.
We have checked that the peaks at $E/\omega=0.6$ and $1.1$ have $J^P=2^-$ and $0^-$, respectively.
The tensor part of the dipole-dipole interaction is again responsible for the splitting.
In the TDHFA the spin dipole mode is unstable and is not shown in Fig. \ref{fig14}.
The spin-orbit mode excited by the operator $({\bm \sigma}\times {\bm r})_z$ which corresponds to $L=1$, $S=1$ and $J^P=1^-$ has zero excitation energy 
at $C=0.8$. We show in Fig. \ref{fig15} the time evolution of $\langle\Phi_0|({\bm \sigma}\times {\bm r})_z|\Phi_0\rangle$ calculated 
in the TDDMA: It is difficult to calculate the strength function because the Fourier transformation requires the TDDMA calculation for quite a long period of time.
The time evolution shows the process toward magnetization induced by the small external field $k({\bm \sigma}\times {\bm r})_z\delta(t-T)$.
Figure \ref{fig15} also shows that the magnetization process is accompanied by small oscillation with frequency $\approx 2\omega$. 
From the above study of the spin excitations in the TDHFA, we can conclude that in a dipolar gas with a spin symmetric and spherical 
closed-shell configuration the instability occurs first in the  $J^P=1^-$ mode followed by
the  $J^P=2^-$ mode and the $J^P=1^+$ mode as a function of $C$. As mentioned above, the difference in the order of unstable modes 
between our result and the result for infinite systems \cite{sogo,li}
is explained by the trapping potential.
The fact that the spin modes calculated in the TDDMA do not show instabilities in the interaction regions where those in the TDHFA do
also suggests that quantum fluctuations (the ground-state correlations and configuration mixing) have an effect of pushing the instabilities to stronger interaction regions.

\section{Summary}
The ground state and collective excitations of an $N=2$ dipolar Fermion gas were studied using the time-dependent density-matrix
approach (TDDMA) which provides us with an alternative way of obtaining the exact solutions. In this approach the physical
observables are directly calculated using the one-body and two-body density matrices
and it has a clear relation to the time-dependent Hartree-Fock theory.
By comparing with the TDDMA results which correspond to the exact solutions we can investigate the effects of quantum fluctuations which are
missing in the mean-field approaches.
It was shown that the magnetization associated with the instability against the Rashba like spin-orbit mode realizes first
and that such magnetization can occur in heavier systems. Comparison with the exact solutions suggests that the instabilities given by 
the Hartree-Fock approximation are shifted to stronger interaction
regimes due to quantum fluctuations.
It was pointed out that the tensor properties of the dipole-dipole interaction can be revealed in the excitations associated with spin degrees of
freedom.
For numerical reasons we were forced to work with rather small configurations spaces and the results
in the TDDMA are not completely converged. We also showed that enlarging the space does not qualitatively
change the results. Therefore, we think that our results are semi-quantitatively correct.
\begin{acknowledgments}
The author would like to thank Dr. P. Schuck for valuable discussions and critical reading of the manuscript.
\end{acknowledgments}

\end{document}